\documentclass{jfm}
\usepackage{graphicx}
\usepackage{epstopdf, epsfig}

\shorttitle{Strongly Stratified Turbulence at Low Prandtl Number}
\shortauthor{V. Skoutnev}

\title{Critical Balance and Scaling of Strongly Stratified Turbulence at Low Prandtl Number}

\author{Valentin A. Skoutnev\aff{1}
  \corresp{\email{valentinskoutnev@gmail.com}}}

\affiliation{\aff{1}Department of Astrophysical Sciences, Princeton University, Princeton, NJ 08544, USA}

\begin{document}

\maketitle

\begin{abstract}
We extend the scaling relations of strongly (stably) stratified turbulence from the geophysical regime of unity Prandtl number to the astrophysical regime of extremely small Prandtl number applicable to stably stratified regions of stars and gas giants. A transition to a new turbulent regime is found to occur when the Prandtl number drops below the inverse of the buoyancy Reynolds number, i.e. $PrRb<1$, which signals a shift of the dominant balance in the buoyancy equation. Application of critical balance arguments then derives new predictions for the anisotropic energy spectrum and dominant balance of the Boussinesq equations in the $PrRb\ll1$ regime. We find that all the standard scaling relations from the unity $Pr$ limit of strongly stratified turbulence simply carry over if the Froude number, $Fr$, is replaced by a modified Froude number, $Fr_M\equiv Fr/(PrRb)^{1/4}$. The geophysical and astrophysical regimes are thus smoothly connected across the $PrRb=1$ transition. Applications to vertical transport in stellar radiative zones and modification to the instability criterion for the small-scale dynamo are discussed.  

\end{abstract}

\section{Introduction}

Turbulence in the strongly stratified regions of planetary oceans, atmospheres and the interiors of stars and gas giants provides an important source of vertical transport of chemicals and momentum, thereby playing a critical role in their long-term evolution \citep{zahn1974rotational,zahn1992circulation,fernando1991turbulent,pinsonneault1997mixing,maeder2000evolution,ivey2008density,ferrari2009ocean,aerts2019angular,garaudJtCS2021}. However, current understanding of the extremely low thermal Prandtl number regime of astrophysical turbulence remains disjoint from the order unity Prandtl number regime of geophysical turbulence. This is despite identical asymptotic limits for the Reynolds $Re\gg1$, Froude $Fr\ll1$, and buoyancy Reynolds $Rb=ReFr^2\gg1$ numbers. The Prandtl number $Pr=\nu/\kappa$ measures the ratio of the microscopic viscosity $\nu$ to the thermal diffusivity $\kappa$ and is extremely small in stellar plasma in particular. Photons rapidly diffuse heat compared to the much slower momentum diffusion by ion-ion collisions, leading to $\kappa\gg\nu$. For reference, stellar radiative zones have Prandtl numbers that can range from $Pr=O(10^{-9})$ to at most $Pr=O(10^{-5})$, in contrast to Earth's fluids that range from  $Pr\simeq 0.7$ in the atmosphere to  $Pr\simeq10$ in the ocean \citep{garaudJtCS2021}. A Prandtl number  as high as $Pr\simeq700$ can be reached in parts of the ocean dominated by salt-stratification  \citep{thorpe2005turbulent,gregg2018mixing,gregg2021ocean} (where the salt diffusivity replaces thermal diffusivity).  Compared to the $Pr=1$ case, a small $Pr$ can have significant effects on large-scale hydrodynamic instabilities and the resulting turbulence \citep{garaudJtCS2021} while a large $Pr$ influences scales comparable to and smaller than the viscous scale and can have significant effects on the buoyancy flux, mixing efficiency, and properties of shear-induced turbulence \citep{salehipour2015turbulent,legaspi2020prandtl,okino2020direct}.

Many important features of the unity Prandtl number regime are becoming better understood from a combination of numerical experiments \citep{waite2004stratified,brethouwer2007,riley2010recent,maffioli2016dynamics,lucas2017layer,de2019effects}, observational data \citep{lindborg2007stratified,riley2008stratified,falder2016seismic, lefauve2022experimental}, and theoretical developments \citep{billant2001self,lindborg2006,chini2022exploiting}. Strongly stratified turbulence with $Pr=O(1)$ forced at some horizontal length scale leads to emergent vertical scales set by the stratification and exhibits two distinct inertial ranges that transfer energy from the large forcing scales to the small viscous and thermal scales where it is dissipated. The transition length scale between the two ranges is known as the Ozmidov scale \citep{ozmidov1992variability}. The turbulence is highly anisotropic from the outer forcing scale down to the Ozmidov scale with the down-scale energy transfer likely containing contributions from both a local energy cascade and non-local energy transfer mechanisms (e.g. shear instabilities, wave-wave interactions) \citep{Waite2011, augier2015stratified,maffioli2016dynamics,khani2018mixing}. Below the Ozmidov scale, the turbulence is isotropic down to the diffusive scales. Properties of the energy cascade, relevant dimensionless parameters, and scaling relations for the emergent vertical scales are now fairly well understood, although many further questions remain such as on the efficiency of mixing \citep{gregg2018mixing,monismith2018mixing,legg2021mixing} and the origin of self-organized criticality \citep{smyth2013marginal,salehipour2018self,smyth2019self,chini2022exploiting, lefauve2022experimental}.    

The main aim of this study is to extend the theoretical arguments used in the $Pr=O(1)$ regime to make analogous predictions for the emergent vertical scales and energy cascades in the asymptotically low $Pr$ regime. If only the thermal diffusivity is increased while keeping all other parameters constant (thereby decreasing $Pr$), one should expect a smooth transition to occur between two asymptotic regimes when thermal diffusion shifts from being important only on the smallest, viscous scales to playing an important role on the mesoscales i.e. on scales comparable to or larger than the Ozmidov scale.

To understand this transition, we use the critical balance framework proposed by \cite{Nazarenko2011} for anisotropic wave systems. Critical balance argues that linear wave $\omega^{-1}$ and non-linear interaction $\tau_{\mathrm{NL}}$ time scales are comparable $\omega^{-1}\sim \tau_{\mathrm{NL}}$ on a scale-by-scale basis throughout a local energy cascade, giving a prediction for the anisotropy of the turbulence as a function of scale. Originally applied in mean field MHD turbulence \citep{goldreich1995toward} (also see \cite{schekochihin2022mhd} and references within), critical balance successfully predicts power laws of the anisotropic energy spectra and associated transition scales in rotating turbulence and unity Prandtl number strongly stratified turbulence \citep{Nazarenko2011}. 

We propose that critical balance should naturally extend to low thermal Prandtl number strongly stratified turbulence with a modification in its physical interpretation. As $Pr$ is decreased (by increasing $\kappa$ while keeping $\nu$ small and fixed), the full internal gravity wave (IGW) dispersion $\omega$ smoothly transitions from the asymptotic dispersion for adiabatic, inviscid, propagating IGWs when $Pr=O(1)$ ($\omega\sim \omega_{\mathrm{N}}$) to an asymptotic dispersion for overdamped, inviscid, IGWs modified by the interaction of buoyancy and fast thermal diffusion when $Pr\ll1$ ($\omega\sim \omega_{\mathrm{lPe}}$, where ``lPe" refers to the ``low turbulent Peclet" limit as discussed in more detail in Sections \ref{sec:ModelAndPrRbTransition} and \ref{sec:CBLowPr}). As a result, scaling laws for the emergent vertical scales and two cascades predicted by critical balance will likewise smoothly change as $Pr$ is decreased. We note that because critical balance assumes a local energy cascade, the relative role of non-local energy transfer mechanisms in the low $Pr$ limit remains to be understood.

\subsection{Paper Layout} 
The Boussinesq equations used to model stably stratified turbulence are defined in Section \ref{sec:ModelAndPrRbTransition} followed by an argument for when a transition of turbulence regimes should occur. Critical balance arguments in the unity $Pr$ regime are reviewed in Section \ref{sec:CBUnityPr} to then compare with the new critical balance arguments in the low $Pr$ regime given in Section \ref{sec:CBLowPr}. Astrophysical applications of the new scaling laws are discussed in Section \ref{sec:Applications}.

\section{The Boussinesq Approximation and the $PrRb=1$ Transition}\label{sec:ModelAndPrRbTransition}
\subsection{The Boussinesq Approximation}
Stably stratified regions of stars and gas giants below their surfaces typically sustain turbulent motions with vertical length scales that are small compared to the local scale height\footnote{The local scale height is the local e-folding scale of a background thermodynamic variable. For example, the pressure scale height in a stellar interior is $H_P\equiv (\partial_r \ln P)^{-1}$.}  and velocity fluctuations are that small compared to the local sound speed. In this limit, the Spiegel-Veronis-Boussinesq equations are a rigorous approximation for the fluctuations of the velocity field and thermodynamic variables of a compressible fluid on top of a stably stratified background \citep{spiegel1960boussinesq}. The approximation effectively filters out the high frequencies of sound waves compared to the lower frequencies of internal gravity waves and fluid motions of interest. The governing equations for the velocity field $\textbf{u}'$ and the buoyancy variable $\theta'=\alpha g T'=-\rho'g/\rho_m$ are:
\begin{subeqnarray}
    \partial_t \textbf{u}'+\textbf{u}'\cdot \nabla \textbf{u}'&=&-\frac{1}{\rho_m}\nabla p'+\theta'\hat{z}+\nu\nabla^2 \textbf{u}',\label{eq:BoussMom}\\
    \partial_t \theta'+\textbf{u}'\cdot\nabla \theta'&=&-N^2u_z'+\kappa\nabla^2\theta',\\
    \nabla \cdot \textbf{u}'&=&0,
\end{subeqnarray}
where $\alpha$ is the coefficient of thermal expansion, $g$ is the local gravitational constant, $T'$ is the temperature perturbation, $\rho'$ is the density perturbation, $\rho_m$ is the mean density of the region, $p'$ is the pressure perturbation, and $N$ is the local Brunt-V\"ais\"al\"a frequency. Primed variables here denote dimensional quantities and unprimed variables will later denote dimensionless quantities.  Note that the Brunt-V\"ais\"al\"a frequency captures all the relevant local thermodynamics of the medium in the Boussinesq approximation of any fluid \citep{bois1991asymptotic}. As a result, the above equations are formally equivalent to those used in geophysical fluid studies of the Earth's oceans, but with a different definition of $N$. For example, in the case of an ideal gas $N^2=\alpha g (\partial_z \bar{T}-\partial_z T_{\rm ad})$ while in the case of a liquid $N^2=-g\partial_z \bar{\rho}/\rho_m$, where $\partial_z\bar{\rho}$, $\partial_z\bar{T}$, and $\partial_z T_{\rm ad}$ are the background density, background temperature, and adiabatic temperature gradients, respectively.

\subsection{Geophysical Regime}\label{sec:GeoPhysRegime}
A stably stratified fluid forced on a horizontal length $L$ and velocity scale $U$ leads to turbulence with an emergent outer vertical length $l_z$ and velocity scale $u_z$ set by the physical parameters of the fluid \{$N$, $\nu$, $\kappa$\}. From dimensional analysis, only three dimensionless parameters characterize the fluid: the Reynolds number $Re=UL/\nu$, the Froude number $Fr=U/NL$, and the Prandtl number $Pr=\nu/\kappa$. Understanding the scaling of the emergent outer vertical scales as well as the structure of the subsequent (anisotropic) energy cascade is an important theoretical and experimental goal. 

In the geophysical fluid regime where $Pr=O(1)$, evidence from theoretical arguments, simulations, and experimental data strongly suggests that the outer vertical length and vertical scales are directly set by the Froude number when the viscosity is sufficiently small: $l_z\sim FrL$ and $u_z\sim FrU$, where $l_z$ is often called the buoyancy length scale. Strong stratification ($Fr\ll1$) thus leads to highly anisotropic structures of the large scale eddies characterized by long horizontal scales and short vertical scales as well as significantly more energy in the horizontal compared to the vertical velocity components. The injected energy undergoes an anisotropic forward energy cascade at large length scales until the Ozmidov scale, an intermediate scale given by $l_O=Fr^{3/2}L$ \citep{brethouwer2007}. For scales smaller than $l_O$, the effects of buoyancy are negligible on the fast turn over times of the eddies and an isotropic Kolmogorov cascade operates down to the dissipation scales, which are the viscous ($l_\nu=Re^{-3/4}L$) and thermal ($l_\kappa\simeq l_\nu$) scales. The range of the isotropic cascade is set by the buoyancy Reynolds number $Rb=ReFr^2$ ($l_O/l_\nu=Rb^{3/4}$) and needs to be sufficiently large in order to support the two cascades characteristic of strongly stratified turbulence \citep{bartello2013sensitivity}.  

The horizontal and vertical scales from above in the limits $Fr\ll1$ and $Rb\gg1$ suggest a consistent rescaling of the Boussinesq equations using the following dimensionalization (identical to \cite{billant2001self}):

\begin{subeqnarray}
\textbf{u}_h'=U \textbf{u}_h,\;u_z'=FrUu_z,\;\theta'=\frac{1}{Fr}\frac{U^2}{L}\theta,\;p'= \rho_m U^2p,\\
x'=Lx,\;y'=L y,\;z'=FrL z,\;t'=\frac{L}{U} t,
\end{subeqnarray}
where the scaling of $\theta'$ is determined by a balance of $\textbf{u}'\cdot\nabla\theta'\sim N^2u_z'$ since thermal diffusivity is considered to be sufficiently small. Substituting the above, the Boussinesq equations become:

\begin{subeqnarray}
    \partial_t \textbf{u}_h+\textbf{u}\cdot \nabla \textbf{u}_h&=&-\nabla_h p+ \left[\frac{1}{Re}\nabla_h^2+\frac{1}{Rb}\nabla_z^2\right]\textbf{u}_h,\label{eq:HorzMomPr1}\\
Fr^2\left[\partial_t u_z+\textbf{u}\cdot \nabla u_z\right]&=&-\nabla_z p+\theta+Fr^2\left[\frac{1}{Re}\nabla_h^2+\frac{1}{Rb}\nabla_z^2\right]u_z,\label{eq:VertMomPr1}\\
    \partial_{t}\theta+\textbf{u}\cdot \nabla \theta&=&-u_z+\left[\frac{1}{PrRe}\nabla_h^2+\frac{1}{PrRb}\nabla_z^2\right]\theta,\label{eq:BuoyPr1}\\
    \nabla\cdot \textbf{u}&=&0,
\end{subeqnarray} 
where $\textbf{u}_h$ and $\nabla_h$ denote the horizontal components of the velocity and gradient. Note that $Rb$ and $PrRb$ act as effective Reynolds numbers in the vertical part of the momentum and thermal diffusion terms, respectively. Examination of the dominant balance to lowest order is helpful:

\begin{subeqnarray}\label{eq:Pr1LowestOrder}
    \partial_t \textbf{u}_h+\textbf{u}\cdot \nabla \textbf{u}_h&=&-\nabla_h p,\label{eq:Pr1horzmom}\\
0&=&-\nabla_z p+\theta,\label{eq:Pr1vertmom}\\
    \partial_t \theta+\textbf{u}\cdot \nabla \theta&=&-u_z,\label{eq:Pr1temp}\\
    \nabla\cdot \textbf{u}&=&0,
\end{subeqnarray} 

Advection in the horizontal momentum equation (Eq \ref{eq:Pr1horzmom}a) is balanced by horizontal pressure gradients, while the dominant balance in the vertical momentum equation (Eq \ref{eq:Pr1vertmom}b) is instead between the vertical pressure gradient and buoyancy fluctuations. These balances will change if the viscosity is increased and the vertical gradients of the momentum diffusion term become important. Further, the buoyancy equation (Eq. \ref{eq:Pr1temp}c) is a balance between temperature advection and displacement of the fluid against the background stratification gradient. It is this latter balance that will change if thermal diffusion is increased and the vertical gradients of the
thermal diffusion term become important, as we show in the next section.

\subsection{Transitions from the Geophysical Regime}\label{sec:TransFromGPR}

We now aim to understand when transitions occur from the regime of geophysical fluid turbulence where $Fr\ll1$, $Rb\gg1$, and $Pr=O(1)$. First, lets consider the better understood transition to the viscosity-affected stratified flow regime as $Re$ is decreased with fixed $Fr$ and $Pr=O(1)$ \citep{godoy2004vertical}. The case of decreasing $Pr$ turns out to behave in an analogous manner. From the perspective of the turbulent cascades, as viscosity is increased, the viscous scale will grow until it is comparable to the Ozmidov scale $l_\nu\sim l_O$, at which point $Rb\sim1$ and the isotropic cascade at small scales disappears \citep{brethouwer2007}. This appears as a shift of the dominant balance in the horizontal momentum equation (Eq \ref{eq:HorzMomPr1}a) between advection and the vertical gradient of the momentum diffusivity:

\begin{equation}
    \frac{\textbf{u}'\cdot \nabla \textbf{u}_h'}{\nu\nabla_z^2\textbf{u}_h'}\sim \frac{u_zl_z}{\nu}\sim ReFr^2=Rb
\end{equation}
where vertical and horizontal advection are comparable due to the incompressibility constraint (i.e. $u_z/l_z\sim U/L$) and $l_z/L\sim u_z/U\sim Fr$ is used to estimate the vertical scales near $Rb\sim1$. Heuristically, $Rb$ is the ratio of the eddy turnover rate to the viscous diffusion rate at the outer vertical scales, i.e. $Rb\sim (u_z/l_z)/(\nu/l_z^2)$. For $Rb<1$,  a change of dominant balance ($\textbf{u}'\cdot \nabla \textbf{u}_h'\sim \nu\nabla_z^2\textbf{u}_h'$) leads to an alternative scaling $l_z/L=Fr/Rb^{1/2}$ (usually written as $l_z/L\sim Re^{-1/2}$) where viscosity dominates the coupling of adjacent vertical layers. The same shift occurs in the buoyancy equation ($N^2 u_z' \sim \kappa\nabla_z^2\theta'$) and the new vertical velocity scale becomes $u_z/U\sim Fr Rb^{1/2}$ (derived with $\theta'\sim \nabla_z p'/\rho_m$ from the unchanged dominant balance in the vertical momentum equation). This regime is often reached by simulations because computational constraints limit how small the viscosity can be set. Note that the vertical scales smoothly transition from $l_z/L\sim u_z/L\sim Fr$ at $Rb=1$ as $Rb$ is decreased.

Returning to the physically interesting limit $Fr\ll1$ and $Rb\gg1$ of strongly stratified turbulence, we now consider the effect of decreasing $Pr$ at fixed $Re$ and $Fr$, equivalent to increasing the thermal diffusivity while keeping the viscosity fixed. From the perspective of the turbulent cascades, as thermal diffusivity is increased, the thermal scale $l_\kappa\sim (PrRe)^{-3/4}L$ will grow until it is comparable to the Ozmidov scale $l_\kappa\sim l_O$, at which point $PrRb\sim1$ and thermal diffusion becomes important on mesoscales that are influenced by buoyancy forces \citep{lignieres2019turbulence}. This appears as a shift of the dominant balance \textit{only} in the buoyancy equation (Eq. \ref{eq:BuoyPr1}c) between advection and the vertical gradient of thermal diffusion:
\begin{equation}
     \frac{\textbf{u}'\cdot \nabla \theta'}{\kappa\nabla_z^2\theta'}\sim \frac{u_zl_z}{\kappa}\equiv Pe_t,
\end{equation}
where $Pe_t$ is the turbulent Peclet number, which can be interpreted heuristically in a similar way to $Rb$ as the ratio of the eddy turnover rate to the thermal diffusion rate at the outer vertical scales, i.e. $Pe_t=(u_z/l_z)/(\kappa/l_z^2)$. If we use the scalings $l_z/L\sim u_z/U\sim Fr$ from the $Pr=O(1)$ regime to estimate $Pe_t$ in the geophysical regime, we see that $Pe_t\sim PrRb$ and so the transition from $Pe_t>1$ to $Pe_t<1$ occurs around $PrRb\sim 1$, exactly like the $l_\kappa\sim l_O$ transition discussed above. Thus, thermal diffusion will cause a transition in turbulent regimes if $PrRb<1$. The scalings for $u_z$ and $l_z$ from the $PrRb>1$ regime will then change (the scaling for $Pe_t$ will correspondingly change as well).

The emergent turbulent Peclet number is a more important parameter than the standard Peclet number $Pe=PrRe$ \citep{zahn1992circulation,lignieres2019turbulence,cope2020dynamics}, which measures the ratio of advection to the horizontal gradient of thermal diffusion: $(\textbf{u}'\cdot \nabla \theta')/(\kappa\nabla_h^2\theta')\sim UL/\kappa$. This is because $Pe_t\ll Pe$---thermal diffusion will always be more important in the vertical than horizontal direction. In astrophysical systems, the extremely large Reynolds numbers often keep $Pe\gg1$ despite small Prandtl numbers. As a canonical example, turbulence from horizontal shear instabilities in the solar tachocline approximately sustain $Re=O(10^{14})$ and $Pe=O(10^{8})$ using $Pr=O(10^{-6})$ \citep{garaud2020horizontal}. Thus horizontal thermal diffusion is likely  less important on outer scales in other stars as well. On the other hand, we find $PrRb=O(10^{1})$ (and hence $Pe_t=O(10^1)$) by using $Fr\sim 3\cdot10^{-4}$ and $Rb\sim O(10^7)$. The near unity value of $Pe_t$ shows that vertical thermal diffusion can easily become relevant in the astrophysical case \citep{garaudJtCS2021}, in particular, in stars with much lower $Pr$ (down to $Pr=O(10^{-9})$ in some stars), stronger background stratification (higher $N$), or weaker driving (lower $U$ or larger $L$). Interestingly, these values of $Re$, $Fr$, and $Rb$ are similar to those found in Earth's atmosphere \citep{lilly1983stratified,waite2014direct}.

An important asymptotic model known as the low Peclet approximation is often used to study the $Pe_t<1$ regime \citep{lignieres1999small}. It can be derived from the shift in dominant balance in the buoyancy equation. If $Pe_t\ll1$, then the $-N^2u_z'$ term balances $\kappa\nabla^2\theta'$ instead of $\textbf{u}'\cdot \nabla \theta'$. In other words, to lowest order, buoyancy fluctuations are generated by vertical advection of the mean background profile while advection of the buoyancy fluctuations is unimportant because of rapid thermal diffusion. As a result, the buoyancy fluctuations can be solved for directly $\theta'=\frac{N^2}{\kappa}\nabla^{-2}u_z'$. This is substituted back into Eq \ref{eq:BoussMom}a to get a closed momentum equation:

\begin{equation}\label{eq:LowPeEqs}
    \partial_t\textbf{u}'+\textbf{u}'\cdot\nabla \textbf{u}'=-\frac{1}{\rho_m}\nabla p'+\frac{N^2}{\kappa}\nabla^{-2}u_z'\hat{z}+\nu\nabla^2 \textbf{u}',\quad \nabla \cdot \textbf{u}'=0
\end{equation}

There are now only two dimensional parameters $N^2/\kappa$ and $\nu$, which, along side the imposed $U$ and $L$, imply that the scaling in this limit can only be functions of the dimensionless parameters $PeFr^{-2}=PrRb/Fr^{4}$ and $Re$ \citep{lignieres2019turbulence,cope2020dynamics}. 

The discussion so far has argued that the geophysical regime with $Fr\ll1$, $Rb\gg1$, and $PrRb>1$ can transition either to a flow dominated by viscous effects when $Rb<1$ or to stratified turbulence modified by thermal diffusion when $PrRb<1$. These transitions correspond to the effective Reynolds numbers of the vertical part of the momentum and thermal diffusion terms becoming smaller than unity, respectively. Our aim is now to derive a formal scaling for the astrophysically motivated $PrRb<1$ regime with the help of the critical balance hypothesis, but first we review critical balance arguments for the geophysical $PrRb>1$ regime.

\section{Critical Balance and Scaling for $PrRb>1$}\label{sec:CBUnityPr}
We derive here the standard scaling relations for the limit of $Fr\ll1$, $Rb\gg1$, and $PrRb>1$ using critical balance arguments. Consider a stably stratified fluid where energy is injected with power $P$ at a low wavenumber $k_f=2\pi/L$ and sustains turbulence with a kinetic energy dissipation rate $\epsilon\sim P\sim  U^3/L$, where $U$ and $L$ are the outer horizontal velocity and length scales. Viewing anisotropic structures in the turbulence as a function of horizontal scale $k_\perp^{-1}$, the goal is to find the characteristic vertical scale $k_\parallel^{-1}$ associated with each $k_\perp$. Here $k_\parallel$ and $k_\perp$ are wavenumbers parallel and perpendicular to the direction of gravity, respectively. Such a structure will have an associated linear timescale $\omega^{-1}$ related to the wave dispersion and a non-linear timescale $\tau_{\rm NL}$ related to the self-straining timescale. We discuss estimation of both timescales for the $PrRb>1$ case before applying critical balance arguments that connect $k_\perp$ and $k_\parallel$.

The Boussinesq system supports linear motions with a dispersion relation given by $\omega(k_\parallel,k_\perp)$.  In the limit of arbitrarily small viscosity and thermal diffusivity, wave damping is negligible  ($\nu k^2,\kappa k^2\ll\omega$) and the linear motions are propagating waves closely approximated by adiabatic, inviscid IGWs with dispersion $\omega\approx\omega_{\mathrm{N}}= N k_\perp/k$. Because large scale vertical motion is strongly restricted, the vertical scales are much finer than the horizontal scales with an anisotropy at large scales quantified by $k_\perp/k_\parallel\ll1$. The linear wave frequency is then approximately:

\begin{equation}
    \omega_{\mathrm{N}}\approx \frac{N k_\perp }{k_\parallel}
\end{equation} 

On the other hand, non-linear interactions breakup eddies and transfer energy from larger to smaller scales, setting up a cascade from the forcing to the diffusive scales where the energy is dissipated. Non-linear interactions occur via the advection term $\textbf{u}\cdot\nabla \textbf{u}$. Incompressibility $\nabla\cdot \textbf{u}=0$ requires $\nabla_\parallel u_\parallel\sim \nabla_\perp u_\perp$ and allows the estimate $\textbf{u}\cdot\nabla \textbf{u}\simeq \textbf{u}_\perp\cdot\nabla \textbf{u}_\perp\simeq \textbf{u}_\parallel\cdot\nabla \textbf{u}_\perp$, where $u_\parallel$ and $u_\perp$ are the scale dependent vertical and horizontal velocities, respectively. The non-linear interaction time scale using the perpendicular non-linearity is given by:
\begin{equation}
    \tau_{\mathrm{NL}}^{-1}\sim k_\perp u_\perp(k_\perp)
\end{equation}

The scaling of $\tau_{\rm NL}$ with $k_\perp$ can be found if a separate relation can connect $u_\perp$ and $\tau_{\rm NL}$. This comes from assuming that a local (in scale) cascade\footnote{Effects of non-local energy transfer mechanisms as well as energy irreversibly lost to the buoyancy flux are not included, but would be needed for a more detailed theory. We expect that the presented arguments should be reasonable as long as the local kinetic energy cascade comprises an $O(1)$ fraction of the total energy cascade.} brings energy down from larger to smaller scales with a ``cascade time" $\tau_{\mathrm{cas}}(k_\perp)$ that determines the energy spectrum $E(k_\perp)$:  

\begin{equation}\label{eq:DefCasTime}
    k_\perp E(k_\perp)\sim u_\perp^2(k_\perp)\sim \epsilon\tau_{\mathrm{cas}}(k_\perp)
\end{equation}
A relation between $\tau_{\rm cas}$ and $\tau_{\rm NL}$ would provide the desired $\tau_{\rm NL}(k_\perp)$. 

In homogeneous, isotropic turbulence the only dimensional option is to set $\tau_{\mathrm{cas}}\sim \tau_{\mathrm{NL}}$, which results in the standard Kolmogorov scalings $u_\perp(k_\perp)\sim(\epsilon/k_\perp)^{1/3}$, $\tau_{\rm NL}\sim (L/U)(k_\perp L)^{-2/3}$, and $E(k_\perp)\sim \epsilon^{2/3}k_\perp^{-5/3}$. In anisotropic turbulence, the two additional dimensionless parameters $k_\parallel/k_\perp$ and $\omega \tau_{\mathrm{NL}}$ no longer constrain the system and the energy spectrum and cascade time can be an arbitrary function of these dimensionless groups. A further physically-motivated constraint is needed.

 \begin{figure}
    \centering
    \includegraphics[width=0.75\linewidth]{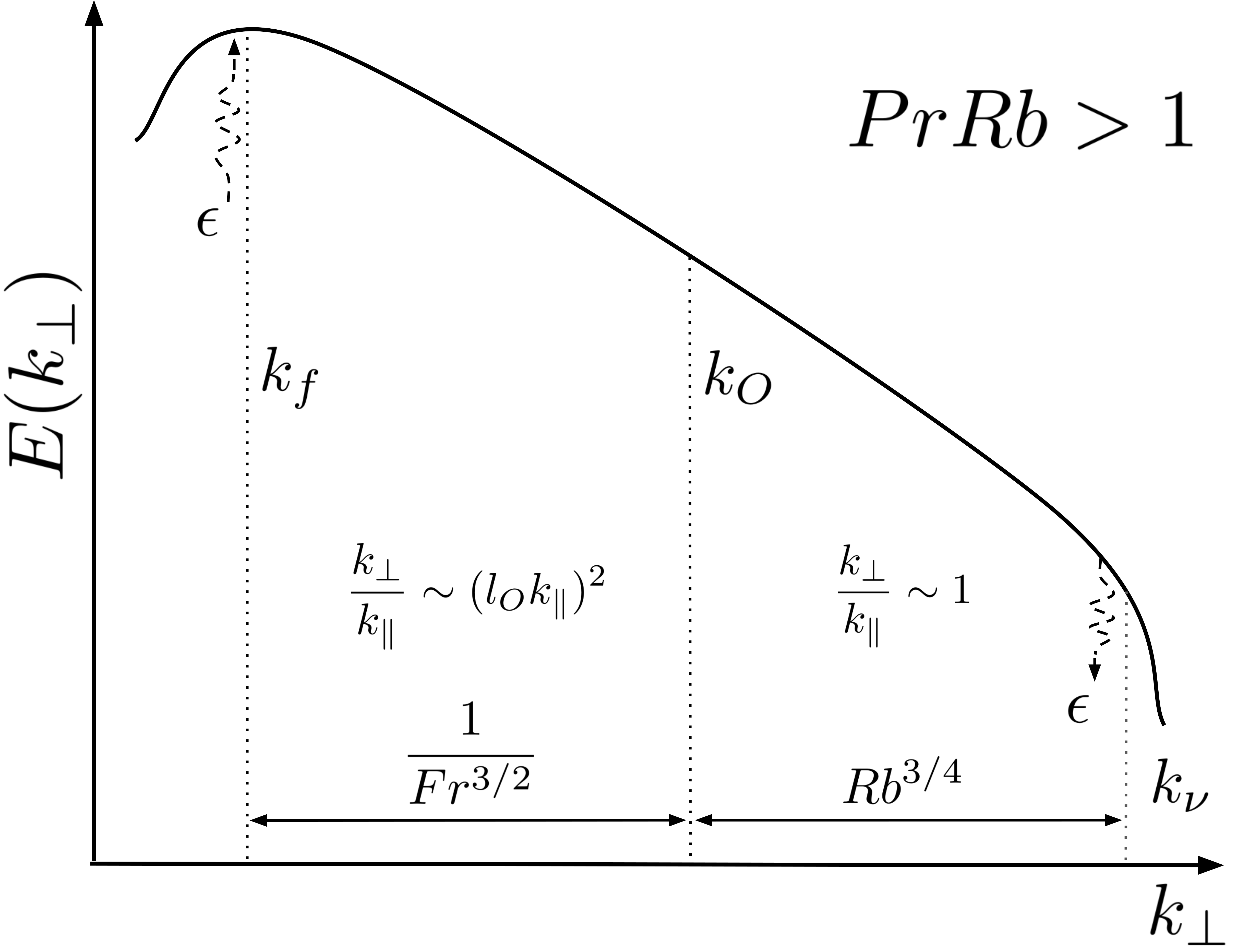}
    \caption{Energy cascade in strongly stratified turbulence for $PrRb\gg1$ relevant to the $Pr=O(1)$ regime of Earth's atmosphere and ocean. $E(k_\perp)$ is the horizontal energy spectrum of the velocity field. Energy is injected at low wavenumber $k_f$ (large scales) with power $P\sim \epsilon$ and undergoes a forward cascade down to dissipation scales. An anisotropic cascade results across horizontal wavenumbers in the range $k_f\lesssim k_\perp\lesssim k_O$ with associated vertical wavenumbers in the range $2\pi/l_z\lesssim k_\parallel\lesssim k_O$. An isotropic cascade follows for wavenumbers larger than $k_O$ up to the dissipation wavenumber $k_\nu$. The strength of the anisotropy is quantified by $k_\perp/k_\parallel$.}
    \label{fig:PrRbGreater1}
\end{figure}

\citet{Nazarenko2011} propose critical balance as a universal scaling conjecture for strong turbulence in anisotropic wave systems. Critical balance states that the linear propagation $\omega^{-1}$ and the non-linear interaction time scales $\tau_{\mathrm{NL}}$ are approximately equal $\omega\tau_{\mathrm{NL}}\sim1$ on a scale by scale basis at all scales where the source of anisotropy is important. Physically, this in essence is a causality argument in a system where the perpendicular non-linearity dominates (e.g. rotating turbulence, MHD with a mean field): a fluctuation with some $k_\perp$ cannot maintain an extent longer than $k_\parallel^{-1}$ set by requiring the linear propagation time in the parallel direction to be comparable to the non-linear breakup time in the horizontal direction. However, the causality argument is more complicated in stably stratified turbulence \citep{Nazarenko2011} because the non-linearity has equal strength in the parallel and perpendicular directions $\tau_{\rm NL}\sim k_\perp u_\perp\sim k_\parallel u_\parallel$, so one could equally argue a balance between linear propagation time in the perpendicular direction and non-linear breakup time in the vertical.  In either case, since the group velocity sets the propagation speed of information, the linear timescale across either the parallel or perpendicular extent of a fluctuation is $ (k_\perp v_{g,\perp})^{-1}\sim(k_\parallel v_{g,\parallel})^{-1}\sim \omega_N^{-1}$. Several physical mechanisms are known to be consistent with critical balance including zigzag \citep{billant2000experimental,billant2000theoretical} and shear instabilities (see Section \ref{sec:VerticalShearInst}), however a complete physical picture is still an area of investigation (see further discussions in \citet{lindborg2006}). Going forward, we assume the critical balance hypothesis and that IGWs effectively set the linear propagation timescale in the $PrRb>1$ limit.

Critical balance removes the ambiguity in determining the cascade time scale (i.e. $\tau_{\mathrm{cas}}\sim\tau_{\mathrm{NL}}$), which results in a Kolmogorov spectrum for the horizontal spectrum at all scales. The vertical spectrum can then be easily determined from $E(k_\parallel)k_\parallel\sim E(k_\perp)k_\perp$ once $k_\perp$ and $k_\parallel$ are related. Applying critical balance $\omega_N \tau_{\rm NL}\sim1$ and rearranging gives the relation between $k_\perp$ and $k_\parallel$:

\begin{equation}\label{eq:Pr1anisotropy}
    k_\perp\sim \left(\frac{\epsilon}{N^3}\right)k^3_\parallel=l_O^2k_\parallel^3
\end{equation}
where $l_O=(\epsilon/N^3)^{1/2}=k_O^{-1}$ is the Ozmidov scale. The anisotropy $k_\perp/k_\parallel\sim (l_Ok_\parallel)^2$ decreases at smaller scales until the Ozmidov scale $k^{-1}\sim l_O$ where the turbulence returns to isotropy $k_\perp\sim k_\parallel$ and $\tau_{\mathrm{NL}}^{-1}\sim N$. The Ozmidov scale is thus the largest horizontal scale that can overturn before restoration by buoyancy forces becomes significant. The horizontal and vertical spectrum for $k<k_O$ are then:

\begin{equation}\label{eq:SpectraScalingHighPr}
    \frac{E(k_\perp)}{U^2L}\sim (k_\perp L)^{-5/3},\quad \frac{E(k_\parallel)}{U^2L}\sim Fr(k_\parallel l_z)^{-3}
\end{equation}
These energy spectra agree with the theoretical scaling predictions in the geophysical literature \citep{dewan1997saturated,billant2001self,lindborg2006}, where the parallel energy spectrum is often written in a dimensional form as $E(k_\parallel)\sim N^2k_\parallel^{-3}$.

The turbulence no longer feels the large scale stratification gradients below the Ozmidov scale since $\omega_{\mathrm{N}} \tau_{\mathrm{NL}}\ll1$ for $k\gg k_O$. Consequently, an isotropic Kolmogorov cascade results at small scales because $\tau_{\mathrm{NL}}$ becomes the only dimensionally available timescale (i.e. the slow buoyancy restoration timescale is irrelevant). The isotropic cascade extends from $k_O$ to the viscous wavenumber $k_\nu$ where $\tau_{\mathrm{NL}}^{-1}\sim \nu k^2\rightarrow l_\nu\sim (\nu/\epsilon^{1/3})^{3/4}$. As a result, the vertical spectrum has break at $k\sim k_O$ from $k_\parallel^{-3}$ to $k_\parallel^{-5/3}$ while the horizontal spectrum remains $k_\perp^{-5/3}$ throughout. The temperature thus plays an important role providing buoyancy for scales $k<k_O$ but is simply advected as a passive scalar for scales $k>k_O$ until it is dissipated at thermal diffusion scales.

For clarity, it is useful to summarize the energy cascade in terms of the dimensionless parameters $Fr$ and $Re$. Energy injected at large horizontal scales $L$ undergoes an anisotropic cascade until the Ozmidov scale, $l_O=Fr^{3/2}L$. Following the anisotropic cascade is an isotropic cascade down to the viscous scale, which can be written as $l_\nu=Re^{-3/4}L$. The size of the isotropic cascade is $l_O/l_\nu=Rb^{3/4}$. $Rb$ essentially plays the role of the effective Reynolds number for the isotropic cascade with outer length scale $l_O$ and velocity scale $u_{\parallel}(k_O)=(\epsilon l_O)^{1/3}=Fr^{1/2}U$ (i.e. $Rb=u_{\parallel}(k_O)l_O/\nu$). A schematic of the energy cascade in terms of dimensionless parameters is shown in Figure \ref{fig:PrRbGreater1}.

Critical balance has provided a prediction for the anisotropy in the energy cascade for $Rb\gg1$ and can therefore also predict the scaling of the outer vertical length and velocity scales of the system by considering the largest scales of the cascade. Substituting  $k_\perp\sim1/L$ and $k_\parallel\sim 1/l_z$ into Eq \ref{eq:Pr1anisotropy} predicts that $l_z\sim FrL$ for the outer vertical length scales. Enforcing incompressibility $u_\parallel\sim (k_\perp/k_\parallel) u_\perp $ subsequently predicts that $u_z=u_\parallel(2\pi/L)\sim FrU$ for the outer vertical velocity scale. These are exactly the scaling relations discussed in Section \ref{sec:GeoPhysRegime}---critical balance successfully reproduces the known scaling relations in the $PrRb>1$ regime.

\section{Critical Balance and Scaling for $PrRb<1$}\label{sec:CBLowPr}

We turn to deriving scaling relations in the $PrRb<1$ regime by extending the critical balance arguments presented in the previous section. The idea is to replace the adiabatic, inviscid IGW frequency $\omega_{\mathrm{N}}$ with the corresponding frequency in the low (turbulent) Peclet limit $\omega_{\mathrm{lPe}}$ for the estimate of the linear timescale. A linear analysis of the Boussinesq equations with $\nu=0$ and $\kappa\neq0$ modeling the $Pr\ll1$ limit gives a dispersion relation:

\begin{equation}\label{eq:omegaLPe}
    \omega^2-i\omega\gamma_\kappa-\omega_{\mathrm{N}}^2=0,\quad \gamma_\kappa=\kappa k^2,\; \omega^2_N=N^2\frac{k^2_\perp}{k^2}
\end{equation}

\begin{figure}
    \centering
    \includegraphics[width=\linewidth]{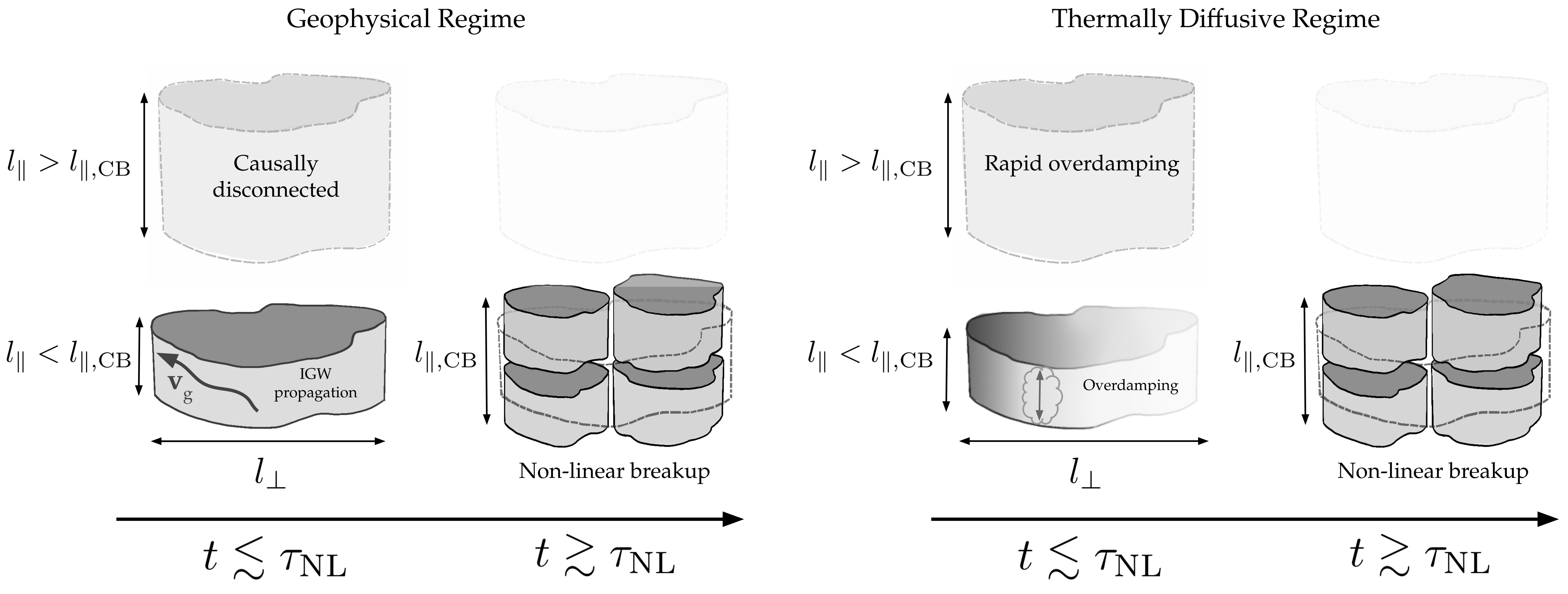}
    \caption{A sketch of the physical arguments used for critical balance for eddies with perpendicular extent $l_\perp\sim k_\perp^{-1}$. The notation $l_{\parallel,\rm CB}$ refers to the $l_\parallel\sim k_\parallel^{-1}$ that satisfies critical balance in each regime. The left side shows the causality argument for the geophysical regime ($PrRb>1$). Since the group velocity, $\textbf{v}_{\rm g}$, of the associated IGW sets the speed at which information can propagate, any eddy with $l_\parallel>l_{\parallel,\rm CB}$ would be causally disconnected before it non-linearly breaks up within time $t\sim \tau_{\rm NL}$. The right side shows the selective decay argument for the thermally diffusive regime ($PrRb<1$). Since the damping rate increases with $\gamma_{\rm lPe}\sim l_\parallel^{4}$, any eddy with $l_\parallel>l_{\parallel,\rm CB}$ would rapidly decay away before it could evolve.  }
    \label{fig:CBSketch}
\end{figure}

In the limit $PrRb\ll1$ ($l_\kappa\gg l_O$), thermal diffusion is faster than the restoring buoyancy timescale so $\gamma_\kappa\gg\omega_{\mathrm{N}}$ and the two roots of Eq \ref{eq:omegaLPe} become $\omega\sim i\omega^2_N/\gamma_\kappa$ and $\omega\sim i\gamma_\kappa$. The latter is an uninteresting rapid thermal diffusion rate, while the former is an effective damping rate $\omega_{\mathrm{lPe}}=i\omega^2_N/\gamma_\kappa=iN^2k_\perp^2/\kappa k^4$ \citep{lignieres1999small,lignieres2019turbulence}. Thus, the linear response frequency is no longer a real frequency of a restoring oscillation, but instead a damping rate $\gamma_{\mathrm{lPe}}$ $(\omega_{\mathrm{lPe}}=i\gamma_{\mathrm{lPe}})$ corresponding to the effective rate that restoring buoyancy and strong thermal diffusion operate on. A direct linear analysis of the low turbulent Peclet equations (Eq \ref{eq:LowPeEqs}) also gives the eponymous damping rate $\omega=i\gamma_{\mathrm{lPe}}$---the two approaches nicely agree. Using the expectation of strong anisotropy at large scales $k_\perp/k_\parallel\ll1$, the linear damping timescale can be estimated as:

\begin{equation}\label{eq:gamma_lPe}
\gamma_{\mathrm{lPe}}\sim \frac{N^2 k_\perp^2}{\kappa k_\parallel^4}
\end{equation}

Before the critical balance hypothesis can be applied, a new justification is needed because the causality argument in the $PrRb>1$ regime no longer applies: waves are overdamped rather than propagate. The instantaneous propagation of information is a peculiarity of the low Peclet equations (Eqs \ref{eq:LowPeEqs}) since the temperature and vertical velocity fields are coupled by an elliptic equation to lowest order. Critical balance instead becomes an argument for selective decay. The dependence between the damping rate of a fluctuation and its vertical extent $\gamma_{\rm lPe}\sim l_\parallel^{4}$ according to Eq \ref{eq:gamma_lPe} means that longer vertical structures overdamp faster. As a result, any fluctuation at some $k_\perp$ with parallel extent longer than the $l_\parallel\sim k_\parallel^{-1}$ set by $\gamma_{\rm lPe}\tau_{\rm NL}\sim1$ will rapidly decay away before non-linear effects can become significant. Critical balance in the thermally diffusive regime thus determines the longest parallel structure for a given $k_\perp$ that can sustain before non-linear breakup. A sketch of the physical argument is shown in Figure \ref{fig:CBSketch}, alongside a comparison with the causality argument in the $PrRb>1$ regime.

With the modified physical interpretation, application of critical balance $\gamma_{\mathrm{lPe}}\tau_{\mathrm{NL}}\sim 1$ and rearrangement gives the new relationship between $k_\perp$ and $k_\parallel$:

\begin{equation}\label{eq:OMAniso}
    \frac{k_\perp}{k_\parallel}\sim \left(\frac{\kappa\epsilon^{1/3}}{N^2}\right)^{3/4}k^2_\parallel=l_{OM}^2k_\parallel^2
\end{equation}
where the modified Ozmidov scale is defined as $l_{OM}=(\kappa\epsilon^{1/3}/N^2)^{3/8}$. One can check that the critical balance prediction self-consistently maintains $\omega_{\mathrm{N}}/\gamma_\kappa\ll1$ all the way to the largest scales since $\omega_{\mathrm{N}}/\gamma_\kappa\sim N k_\perp/\kappa k_\parallel^3\sim (PrRb)^{1/4}\ll1$. The turbulence now returns to isotropy at the modified Ozmidov scale where the overdamping rate for a fluctuation with $k_\perp\sim k_\parallel$ is comparable to its eddy turnover time $N^2l_{\rm OM}^2/\kappa\sim \tau_{\rm NL}^{-1}$. In analogy to the Ozmidov scale, the modified Ozmidov scale is the largest horizontal scale that can overturn before overdamping becomes significant. Note that the modified Ozmidov is larger than the Ozmidov scale as would be expected:

\begin{equation}
    l_{OM}=\frac{Fr^{3/2}}{(PrRb)^{3/8}}L,\quad \frac{l_{OM}}{l_O}=(PrRb)^{-3/8}>1
\end{equation}

\begin{figure}
    \centering
    \includegraphics[width=0.75\linewidth]{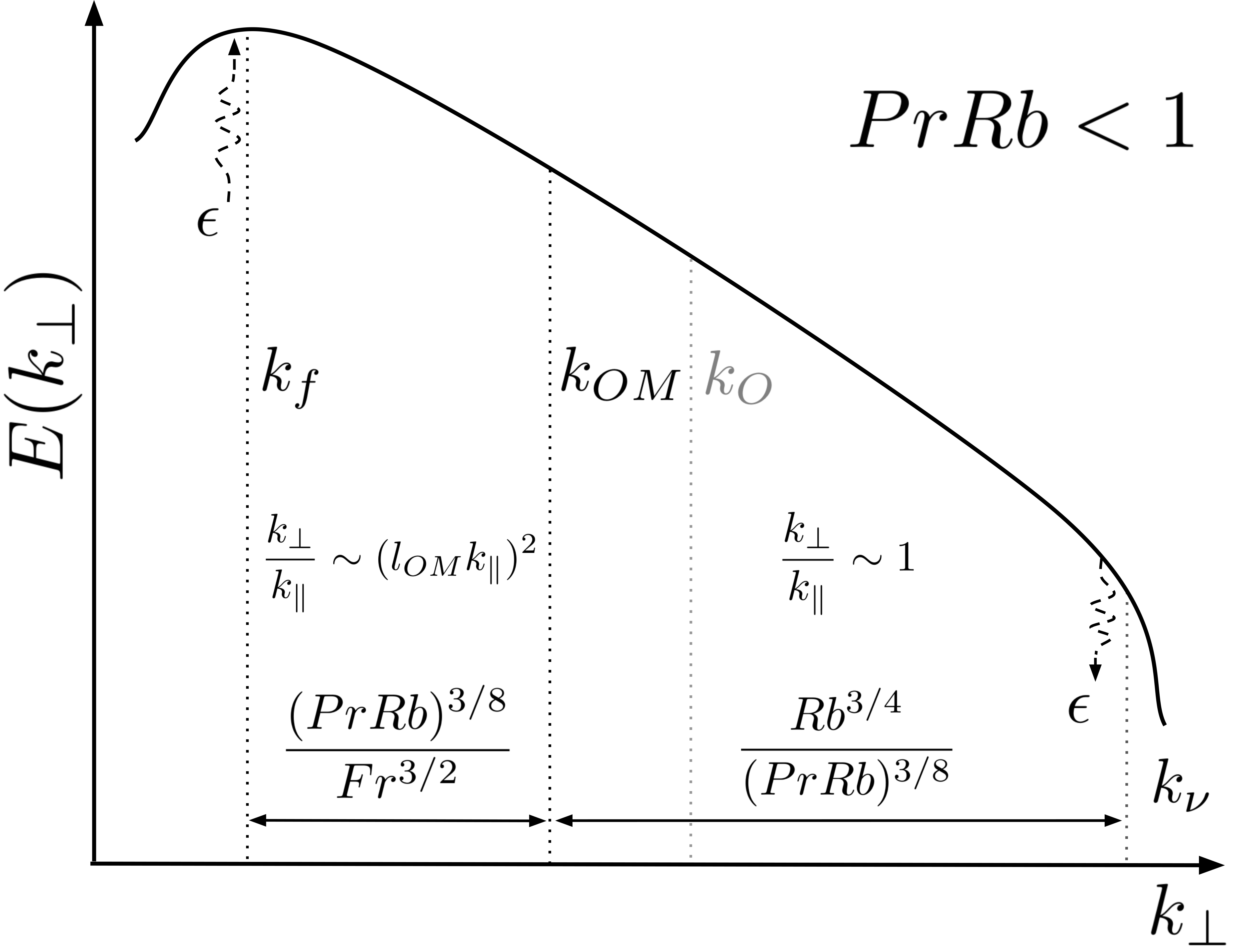}
    \caption{Energy cascade for strongly stratified turbulence for $PrRb\ll1$ relevant to the $Pr\ll1$ regime of stellar radiative zones. Similar to Figure \ref{fig:PrRbGreater1}, an anisotropic cascade results in the range $k_f\lesssim k_\perp \lesssim k_{OM}$ followed by an isotropic cascade from $k_{OM}$ to $k_\nu$. Note that the beginning of the isotropic subrange has moved to larger scales $k_{OM}<k_O$.}
    \label{fig:PrRbLess1}
\end{figure}

The outer vertical scale ($k_\parallel\sim l_z^{-1}$) at the largest horizontal scale where $k_\perp\sim L^{-1}$ can again be found. Subsequent enforcement of incompressibility $(\nabla\cdot \textbf{u}'=0)$ then gives $u_z\sim l_z U/L$. The result is shown below:

\begin{equation}
    l_z\sim \frac{Fr}{(PrRb)^{1/4}}L, \quad u_z\sim \frac{Fr}{(PrRb)^{1/4}}U.
\end{equation}

These scalings self-consistently predict a small turbulent Peclet number $Pe_t\sim (PrRb)^{1/2}\ll1$. At this point it becomes suggestive to define a modified Froude number as $Fr_M\equiv Fr/(PrRb)^{1/4}$ so that:

\begin{equation}
    l_z=Fr_ML,\quad l_{OM}=Fr_M^{3/2}L.
\end{equation}
These are the same exponents as in the geophysical regime. Using the new definition, the corresponding horizontal and vertical spectra for $k<k_{OM}$ are: 

\begin{equation}\label{eq:SpectraScalingLowPr}
    \frac{E(k_\perp)}{U^2L}\sim (k_\perp L)^{-5/3},\quad \frac{E(k_\parallel)}{U^2L}\sim Fr_M(k_\parallel l_z)^{-3}
\end{equation}

\begin{figure}
    \centering
    \includegraphics[width=0.75\linewidth]{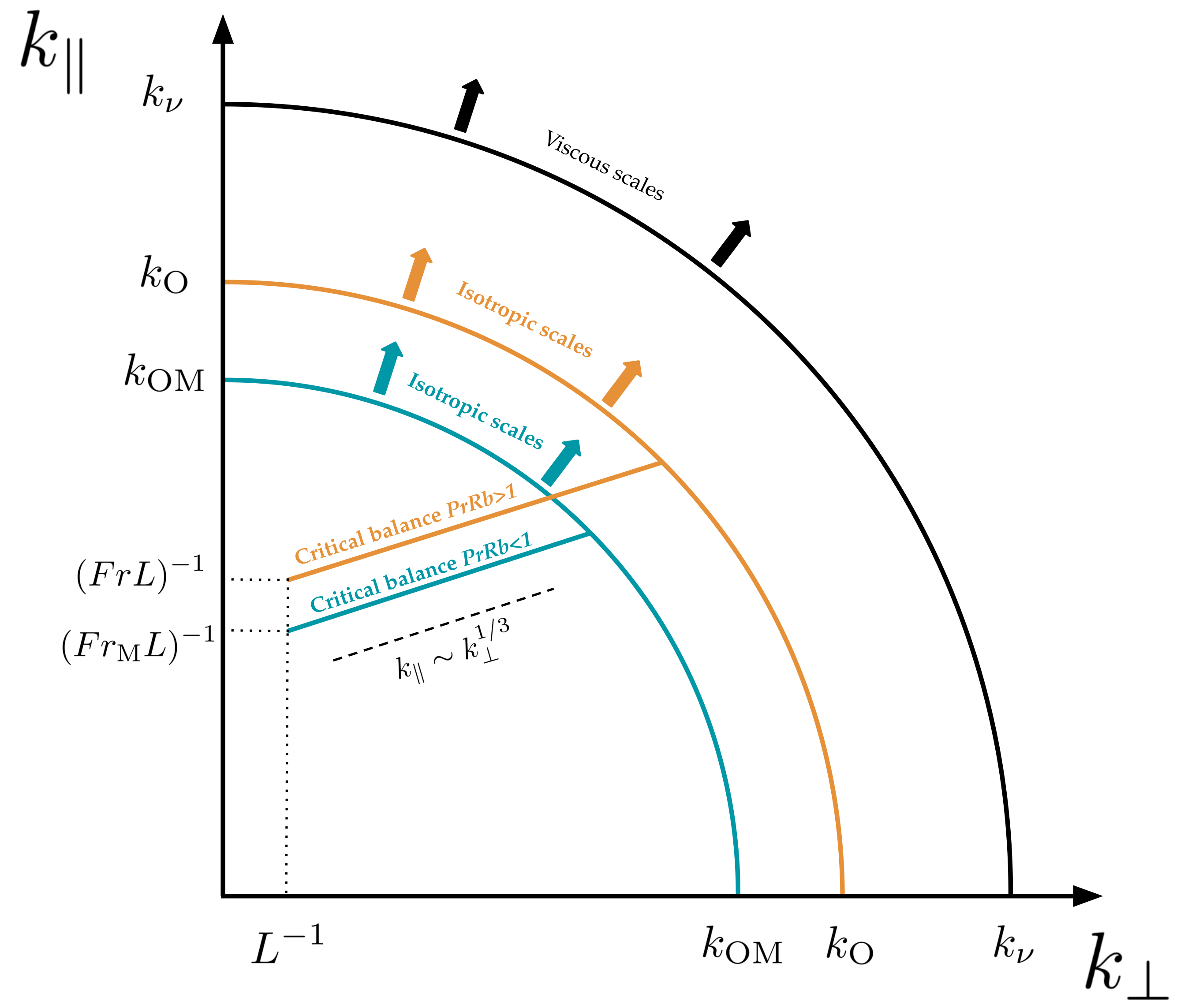}
    \caption{Comparison of the cascade path in the $k_\perp-k_\parallel$ plane between the geophysical (teal) and thermally diffusive (orange) regimes. Energy is injected at low perpendicular wavenumber $k_\perp\sim L^{-1}$, follows the critical balance path until the respective Ozmidov scale ($k_{\rm O}$ or $k_{\rm OM}$), enters the isotropic cascade, and then is dissipated at the viscous scales.  }
    \label{fig:PerpParPlane}
\end{figure}

The dimensional form of the parallel energy spectrum $E(k_\parallel)\sim N(\epsilon/\kappa)^{1/2}k_\parallel^{-3}$ now depends on the dissipation and thermal diffusivity, unlike in the geophysical regime where $E(k_\parallel)\sim N^2k_\parallel^{-3}$. A schematic of the energy cascade is shown in Figure \ref{fig:PrRbLess1} and a comparison of the cascade path in the $k_\perp-k_\parallel$ plane with the $PrRb>1$ regime is shown in Figure \ref{fig:PerpParPlane}.

It is now clear that the transition from the $PrRb>1$ to the $PrRb<1$ regime simply corresponds to a replacement $Fr\rightarrow Fr_M$. How can this modified Froude number be physically understood? If the Froude number is reinterpreted to \textit{define} the ratio of the emergent vertical length scale to the imposed horizontal scale (i.e. $l_z/L\equiv Fr$), then critical balance at the largest scales simply sets the Froude number. By substituting $k_\perp\sim L^{-1}$ and $k_\parallel\sim l_z^{-1}$ into the linear wave frequencies $\omega_{\mathrm{N}}\sim N l_z/L$ and $\gamma_{\mathrm{lPe}}\sim N^2l_z^4/\kappa L^2$ and then comparing both with the corresponding non-linear frequency scale $\tau_{\mathrm{NL}}^{-1}\sim U/L$, the two Froude numbers emerge:
\begin{subeqnarray}
    PrRb>1&:&\quad \frac{l_z}{L}\sim \frac{U}{LN}\equiv Fr, \\  
    PrRb<1&:&\quad \frac{l_z}{L}\sim \left(\frac{\kappa U}{N^2L^3}\right)^{1/4}\equiv Fr_M.
\end{subeqnarray}

As should be expected, the outer vertical scales smoothly transition from $l_z/L=Fr$ to $l_z/L=Fr/(PrRb)^{1/4}$ at $PrRb=1$ when $Pr$ is decreased. This is analogous to the smooth transition from $l_z/L=Fr$ to $l_z/L=Fr/(Rb)^{1/2}$ at $Rb=1$ when $Re$ is decreased, as discussed in Section \ref{sec:TransFromGPR}. Additionally, we note that $Fr_M$ can be rewritten in terms of $PeFr^{-2}$ and $Re$ (i.e. $Fr_M=(PeFr^{-2})^{-1/4}$) as was required by the low turbulent Peclet approximation.

With the acquired scaling relations from critical balance above, we suggest the following dimensionalization for rescaling the Boussinesq equations in the $PrRb\ll1$ limit:

\begin{subeqnarray}
\textbf{u}_h'=U \textbf{u}_h,\;u_z'=Fr_MUu_z,\;\theta'=\frac{1}{Fr_M}\frac{U^2}{L}\theta,\;p'= \rho_m U^2p,\\
x'=Lx,\;y'=L y,\;z'=Fr_ML z,\;t'=\frac{L}{U} t,
\end{subeqnarray}
Aside from $z'$ and $u_z'$, the only other variable whose scaling changed is $\theta'$, which is now determined by the new dominant balance between $N^2u_z'\sim\kappa \nabla_z^2\theta'$ as discussed in Section \ref{sec:TransFromGPR}. The Boussinesq equations become:

\begin{subeqnarray}
    \partial_t \textbf{u}_h+\textbf{u}\cdot \nabla \textbf{u}_h&=&-\nabla_h p+ \left[\frac{1}{Re}\nabla_h^2+\frac{1}{Rb_M}\nabla_z^2\right]\textbf{u}_h,\\
Fr_M^2\left[\partial_t u_z+\textbf{u}\cdot \nabla u_z\right]&=&-\nabla_z p+\theta+Fr_{M}^2\left[\frac{1}{Re}\nabla_h^2+\frac{1}{Rb_M}\nabla_z^2\right]u_z,\label{eq:vertMomLPE}\\
    (PrRb)^{1/2}\left[\partial_{t}\theta+\textbf{u}\cdot \nabla \theta\right]&=&-u_z+ \left[\frac{(PrRb)^{1/2}}{PrRe}\nabla_h^2+\nabla_z^2\right]\theta,\\
    \nabla\cdot \textbf{u}&=&0,
\end{subeqnarray} 

We see that $Rb_M\equiv ReFr_M^2=Rb/(PrRb)^{1/2}$ acts as the new effective Reynolds number and nicely matches with the scale separation between the modified Ozmidov and the viscous scale $l_{OM}/l_\nu=(Rb_M)^{3/4}$. To lowest order:

\begin{subeqnarray}
    \partial_t \textbf{u}_h+\textbf{u}\cdot \nabla \textbf{u}_h&=&-\nabla_h p,\\
0&=&-\nabla_z p+\theta,\\
    0&=&-u_z+\nabla_z^2\theta,\\
    \nabla\cdot \textbf{u}&=&0,
\end{subeqnarray} 

Comparing with the $PrRb>1$ Equation set \ref{eq:Pr1LowestOrder}, the only difference appears in the dominant balance of the buoyancy equation where $u_z\sim\nabla_z^2\theta$ instead of $u_z\sim \textbf{u}\cdot 
\nabla\theta$, as expected. The vertical momentum equation remains a balance of buoyancy fluctuations with the vertical pressure gradient. Advection of the vertical momentum is now suppressed by a factor of $Fr_M^2$ instead of $Fr^2$. In principle, this term can become order unity at sufficiently low $Pr$ if $Pr<Fr^2/Re$. Equivalently, when thermal diffusion is so efficient that $k_{OM}<k_f$  (or $\kappa>N^2L^3/U$), at which point the turbulence is simply isotropic.

Looking back at all the arguments so far in both the $PrRb<1$ and $PrRb>1$ regime, we note the parallelism between the critical balance arguments for $\omega\tau_{\mathrm{NL}}$ and the dominant balance arguments for the Boussinesq equations. $\tau_{\mathrm{NL}}$ represented the non-linear advection terms (i.e. $\textbf{u}\cdot \nabla \textbf{u}_h$) while $\omega$ captured the combined effects of all the remaining linear terms. As dominant balance shifted between linear terms in various equations, $\omega$ shifted correspondingly to the dominant root of the full IGW dispersion relation (and visa versa). For example, $\omega_{\mathrm{N}}\rightarrow i\gamma_\nu\equiv i\nu k_z^2$ as $Re$ was decreased in Section \ref{sec:ModelAndPrRbTransition} and $\omega_{\mathrm{N}}\rightarrow \omega_{\mathrm{lPe}}$ as $Pr$ was decreased in Section \ref{sec:CBLowPr}. Critical balance and the corresponding dominant balance arguments are therefore one and the same.

\subsection{Role of Vertical Shear Instabilities}\label{sec:VerticalShearInst}

Examining the role of vertical shear instabilities in the turbulence offers an alternative physical interpretation of the scaling results. Many experiment and numerical studies in the geophysical regime find evidence suggesting that turbulent structures are organized to be marginally unstable to local vertical shear instabilities, a behaviour thought to be closely related to self-organized criticality \citep{smyth2013marginal,salehipour2018self,smyth2019self,chini2022exploiting,lefauve2022experimental}. A heuristic argument for such behavior is that the vertical shear of a turbulent eddy will generally be driven towards the marginal value of the shear for instability. An eddy with a strong vertical shear will go unstable within a dynamical time and reduce its shear to the marginal value, while one with a weak shear can become amplified to the marginal value before going unstable. As a result, vertical gradients of the horizontal velocity of eddies may be maintained at the marginal condition for vertical shear instability. As a concrete example, vertically adjacent quasi-horizontal eddy motions
(e.g. ``pancake" eddies or modes) have been suggested and observed to exhibit such behaviour in early studies such as \cite{lin1979wakes,lilly1983stratified, spedding1996long,riley2003dynamics} (and references within). We now examine how the critical balance scalings in both the geophysical and thermally diffusive regimes marginally satisfy the corresponding vertical shear instability criteria.     

In the $PrRb>1$ ($Pe_t\sim PrRb>1$) regime, the non-linear criterion to sustain turbulence in a vertical shear instability requires $J=N^2/S^2\lesssim 1$ \citep{richardson1920supply,howard1973stability}  (i.e. Richardson's criterion), which is essentially based on an energy argument requiring more kinetic energy release than potential energy cost upon instability of the flow. To satisfy marginal stability, eddies with horizontal velocity scale $u_\perp(k_\perp)$ would need to maintain a vertical separation $l_\parallel$ such that $J(k_\perp)=N^2/S^2\sim 1$ holds with $S\sim u_\perp/l_\parallel$. Indeed, substituting the critical balance results ($k_\perp E(k_\perp)\sim u_\perp^2(k_\perp)$ alongside Eq. \ref{eq:Pr1anisotropy}) into $J(k_\perp)\sim N^2l_\parallel^2/u_\perp^2(k_\perp)$ shows that $J(k_\perp)\sim 1$ is satisfied on scale-by-scale basis from $k_f\leq k_\perp\leq k_O$. Interpretation of strongly stratified turbulence in terms of marginal instability of local vertical shear instabilities is therefore consistent with critical balance scalings in the geophysical regime.

In the $PrRb<1$ ($Pe_t\sim (PrRb)^{1/2}<1$) regime, the non-linear instability criterion is relaxed due to the weakening of the potential energy cost by the significant thermal diffusion rate at large vertical scales ($u_z/l_z<\kappa/l_z^2$ since $Pe_t<1$). The criterion proposed on phenomenological grounds by \cite{zahn1992circulation} requires $J Pe_t\lesssim 1$ for instability, which smoothly connects across the $Pe_t=1$ transition and has been approximately verified (in accessible parameter regimes) in recent numerical studies \citep{prat2013turbulent,prat2014shear,garaud2015stability,prat2016shear,garaud2017turbulent}. To satisfy marginal stability, eddies with horizontal velocity scale $u_\perp(k_\perp)$ would need to maintain a vertical separation $l_\parallel$ such that $J(k_\perp)Pe_t(k_\perp)=N^2u_\parallel l_\parallel/ S^2\kappa\sim 1$. Similar to before, substituting the critical balance results ($k_\perp E(k_\perp)\sim u_\perp^2(k_\perp)$ alongside Eq. \ref{eq:OMAniso}) into $J(k_\perp)Pe_t(k_\perp)$ shows that $J(k_\perp)Pe_t(k_\perp)\sim 1$ is satisfied on a scale-by-scale basis from $k_f\leq k_\perp\leq k_{\rm OM}$. Thus the consistent interpretation between the marginal instability of local vertical shear instabilities and critical balance scalings is maintained across the $Pe_t=1$ transition into the thermally diffusive regime.

We note that some studies add the additional requirement $Re_t=u_zl_z/\nu>Re_{\rm crit}\sim 10^3$ to ensure that viscous effects play no role in suppressing the non-linear criterion for instability. Combining this assumption with $JPe_t\lesssim 1$ gives a weaker condition for instability $JPr\lesssim (JPr)_{\rm crit}$  (necessary but not sufficient since this is equivalent to $JPe_t\lesssim Re_t/Re_{\rm crit}$). Reporting simulations in terms of $JPr$ can useful since it may be a proxy of $Re_t\sim (JPr)^{-1}$ \citep{prat2016shear}. However, requiring $Re_t\gg1$ as an additional assumption is unnecessary for strongly stratified turbulence because $Re_t\sim Rb_M\gg1$ is already built into the asymptotics. This is physically reasonable because the dynamics of scales larger than the modified Ozmidov scale cannot be affected by viscous effects (i.e. since $l_\nu\ll l_{\rm OM}$).

\subsection{Comparison with Previous Work}\label{sec:CompWCope}
This work follows in the light of a series of simulations and an evolving discussion in \cite{cope2020dynamics} and \cite{garaud2020horizontal} attempting to understand the complicated parameter space of $Pr<1$ stratified turbulence driven by horizontal shear instabilities. \cite{cope2020dynamics} examines the low Peclet limit where $Pe<1$ and $Pe_t\ll1$, which falls squarely into our $PrRb<1$ regime and allows for comparison. Predictions of the vertical outer scales from critical balance ($u_z\sim Fr_MU$, $l_z\sim Fr_ML$) are in conflict with the theoretical predictions of \cite{cope2020dynamics} shown below:

\begin{equation}\label{eq:Cp2020Scalings}
u_z\sim Fr_M^{2/3}U,\quad l_z\sim Fr_M^{4/3}L,    
\end{equation}
where their notation has been translated using $Fr_M=(BPe)^{-1/4}$.

A major source of the discrepancy arises due to different assumptions in \cite{cope2020dynamics} for the dominant balance of the vertical momentum equation. Examining the proposed balances directly on the low turbulent Peclet equations gives a clear way to compare. The horizontal component of the momentum equation (Eq. \ref{eq:LowPeEqs}) requires the pressure to be order unity $p\sim U^2/L$. For the vertical component, our arguments in this work essentially balance $\nabla_z p'\sim (N^2/\kappa)\nabla^{-2}u_z'$, which directly implies $l_z\sim Fr_M L$ by using $\nabla^{-2}\sim l_z^2$ and enforcing incompressibility $u_z/l_z\sim U/L$ at the outer scales. The vertical advection terms are then smaller by $O(Fr_M^2)$ (see Eq. \ref{eq:vertMomLPE}b) and result in a consistent balance similar to the $PrRb>1$ regime (see Eq \ref{eq:VertMomPr1}b).  \cite{cope2020dynamics} alternatively assume a balance of vertical advection and the thermally-modified buoyancy term: $\textbf{u}'\cdot \nabla u_z'\sim  (N^2/\kappa)\nabla^{-2}u_z'$. This assumption leads to an inconsistency because it results in a vertical pressure gradient that is too large ($\nabla_z p'\gg \textbf{u}'\cdot \nabla u_z'$), as discussed in Section 5.2.2 of \cite{cope2020dynamics}. It is unclear how Eq. \ref{eq:Cp2020Scalings} would smoothly transition from the $PrRb>1$ regime, perhaps requiring an additional intermediate turbulence regime. We also note that Eqs \ref{eq:Cp2020Scalings} do not satisfy the incompressibility constraint, which stems from their assumption that horizontal and vertical lengths scales of the spatial derivatives are similar. This assumption may be justified in their case where turbulence is driven in a wide band of horizontal modes by the horizontal shear instability, while our arguments suppose an idealized forcing at a single horizontal scale.

The large set of $Pe<1$ simulations in \cite{cope2020dynamics} (e.g. Figure 8) should allow differentiation between the two theoretical predictions in principle: however, the comparison is difficult because of the small differences between the fractional exponents. A detailed follow up analysis of the simulations in light of the new critical balance predictions hopefully should resolve the question of the correct scaling relations and will be explored in forthcoming work. It is promising that \cite{garaud2020horizontal} find a sharp transition at $Pe_t\sim1$ as shown in their Figure 3, which we interpret as the $PrRb\sim1$ transition. 

We note that \cite{garaud2020horizontal} predicts an additional turbulent regime for $Pr<1$, $Pe\gg1$, and $Pe_t>1$. According to the critical balance framework, this falls into the $PrRb>1$ regime and scaling from the geophysical literature ($l_z\sim FrL$) should simply apply. \cite{garaud2020horizontal} instead theoretically propose and find numerical evidence for scaling relations given by $u_z/U\sim l_z/L\sim Fr^{2/3}$. Similar to our $PrRb>1$ regime, \cite{garaud2020horizontal} argues $\nabla_zp'\sim\theta'$ and $\textbf{u}'\cdot \nabla\theta'\sim-N^2u_z'$, but instead of assuming incompressiblity to get $l_z/L\sim Fr$, they argue for a constant scaling of the time-dependent mixing efficiency to arrive at $l_z/L\sim Fr^{2/3}$ with motivation and support from their numerical results (see their Figure 4). As discussed in Section 4.4.4 of \cite{garaud2020horizontal}, several factors may be at play. First, $l_z$ may be sensitive to the method used to extract the vertical scale from simulations. For example, \cite{lindborg2006} use a weighed average of $k_z$ with the energy spectrum to find a $l_z/L\sim Fr$ scaling in $Pr=1$ simulations while \cite{garaud2020horizontal} use the vertical correlation length in their $Pr=O(10^{-1})$ simulations. The different methods correspond to different weighting functions when integrating with the energy spectrum. It would be useful to compare both diagnostics at the same $Pr$ to resolve this possible issue. Another issue is that reaching asymptotic values of $Rb>>1$ while keeping $Fr\ll1$ is computationally expensive due to a requirement for two large scale separations between $L$ and $l_{\rm OM}$ as well as $l_{\rm OM}$ and $l_{\nu}$. Measuring scaling exponents in simulations with $Fr=O(10^{-1})$ and $Rb=O(10^1)$ may lead to transitional scaling relations \citep{bartello2013sensitivity}. Indeed, the arguments presented in this paper and the theoretical literature, strictly speaking, only rigorously hold for the asymptotic limits of $PrRb\gg1$ or $PrRb\ll1$.

\section{Astrophysical Applications}\label{sec:Applications}

\subsection{Diffusion Coefficients}
Mixing of chemical elements in stellar interiors by stratified turbulence can have important consequences for stellar evolution and quantifying its efficiency is essential for comparison with stellar observations (see \cite{maeder2000evolution,salaris2017chemical} and references within). 1D stellar evolution models require an effective vertical diffusion coefficient, $D_v$, that is typically estimated by the product of some characteristic vertical velocity and length scales of the turbulence, i.e. $D_v\sim u_z l_z$. Which scales to choose remains an important problem and may depend on the instability that drives the turbulence. Typical choices include using the outer vertical scales or the largest isotropic scales. The scale-dependent anisotropy obtained from critical balance allows an estimate of the contribution to $D_v$ from eddies of different scales in stratified turbulence with horizontal length scale $L$ and vertical scale $U$ set by some instability (e.g. shear instabilities of differential rotation). We define the scale-dependent diffusion coefficient $\tilde{D}_v(k_\parallel)=u_\parallel(k_\parallel)k_\parallel^{-1}$ from eddies of vertical length scale $k_\parallel^{-1}$ and corresponding vertical velocity scale $u_\parallel(k_\parallel)$. Using the incompressibility relation and the anisotropy relations for $k_\perp/k_\parallel$ gives:

\begin{subeqnarray}
    k_\parallel\leq k_{O(M)}&:&\quad \frac{\tilde{D}_v(k_\parallel)}{UL}=Fr_{(M)}^2\\
    k\geq k_{O(M)}&:&\quad \frac{\tilde{D}_v(k)}{UL}=\frac{Fr_{(M)}^2}{(kl_{O(M)})^{4/3}}\quad (k_\parallel\sim k)
\end{subeqnarray}
where $k_{O(M)}$ is defined to be $k_O$ for $PrRb>1$ or $k_{OM}$ for $PrRb<1$ (and similarly for $l_{O(M)}$ and $Fr_{(M)}$).
The turbulent diffusion coefficient in principle has contribution from turbulence at all scales, but is often argued to be dominated by either the outer scales $D_v\sim \tilde{D}_v(l_z^{-1})$ or the Ozmidov scales $D_v\sim \tilde{D}_v(l_{O(M)}^{-1})$. The constant value of $\tilde{D}_v$ in the large-scale anisotropic subrange for $k_\parallel<k_{O(M)}$ means that one can equally use either choice, i.e. $D_v\sim u_\parallel(l_z^{-1})l_z\sim u_\parallel(l_{O(M)}^{-1})l_{O(M)}$. Thus scaling relations based on critical balance predict that $D_v\sim Fr^2UL$ in the $PrRb>1$ regime and $D_v\sim Fr_M^2UL$ in the $PrRb<1$ regime. The estimated turbulent diffusion coefficient in the $PrRb<1$ limit is larger than in the $PrRb>1$ limit by a factor of:  

\begin{equation}
    \frac{D_v^{[PrRb<1]}}{D_v^{[PrRb>1]}}\sim \left(\frac{Fr_M}{Fr}\right)^2=\frac{1}{(PrRb)^{1/2}}
\end{equation}
which is significant only if $PrRb\ll1$. 

The foundational work of \cite{zahn1992circulation} (also see \cite{lignieres2019turbulence}) use the modified Ozmidov scales $D_v\sim u_\parallel(l_{OM}^{-1})l_{OM}\sim (\epsilon \kappa/N^2)^{1/2}$, in agreement with the prediction above. Our estimates for $D_v$ therefore do not affect the results of previous astrophysical studies based on the estimate for $D_v$ by \cite{zahn1992circulation}. Lastly, we note that the diffusion coefficient in the $PrRb<1$ limit agrees with the prediction from \cite{cope2020dynamics} (using the outer vertical scales for estimation of $D_v\sim u_\parallel(l_z^{-1})l_z$), although the agreement may be coincidental given their different predictions for $u_z$ and $l_z$ as discussed in Section \ref{sec:CompWCope}. 

\subsection{Small-scale Dynamo Instability Criterion}
Turbulence in the fully ionized and highly conductive plasma of a stellar radiative zone may be able to generate and sustain magnetic fields on scales smaller than the forcing scale through the small-scale dynamo (SSD). Anisotropy in the velocity field caused by stable stratification makes dynamo action less efficient compared to isotropic turbulence at the same magnetic Reynolds number $Rm=UL/\eta$, where $\eta$ is the resistivity. Indeed, the limit of infinitely strong stratification leads to a planar velocity field (i.e. $\textbf{u}=(u_x(x,y,z,t),u_y(x,y,z,t),0)$), which is a known anti-dynamo flow \citep{zeldovich1980magnetic}. Using a large set of direct numerical simulations, \cite{skoutnev2021small} found that the SSD is unstable in the $Pr=O(1)$ regime if the scale separation of the Ozmidov scale and the magnetic resistive scale is sufficiently large. Quantitatively, this translates to requiring the magnetic buoyancy Reynolds number $Rb_m=PmRb$ to be larger than a critical value (i.e. $Rb_m>Rb_m^c$), where $Pm=\nu/\eta$ is the magnetic Prandtl number. The direct analogy to isotropic turbulence is the requirement that $Rm$  be larger than a critical value $Rm>Rm^c$. 

We propose that in the $PrRb<1$ regime, the SSD criterion switches to requiring a sufficient scale separation between the modified Ozmidov scale and the resistive scale. The criterion in the two regimes is then given by:
\begin{subeqnarray}
    PrRb>1&:&\quad Rb_m=PmRb>Rb_m^c\\
    PrRb<1&:&\quad Rb_{m,M}=PmRb_M=\frac{PmRb}{(PrRb)^{1/2}}>Rb_{m,M}^c
\end{subeqnarray}

This extension naturally carries through the conjecture in \cite{skoutnev2021small} that the SSD only operates within the isotropic portion of the turbulent cascade. The new criterion predicts that the SSD will become more unstable as $Pr$ is decreased at fixed $Rb$ into the $PrRb<1$ regime since $Rb_M>Rb$. Qualitatively, dynamo action becomes more efficient as the anisotropy caused by stratification is decreased by increased thermal diffusion.

We expect that both $Rb_m^c=Rb_m^c(Pm)$ and $Rb_{m,M}^c=Rb_{m,M}^c(Pm)$ should have a weak $Pm$ dependence that is strongest around $Pm=O(1)$, similar to the dependence of $Rm^c(Pm)$ in the case of isotropic turbulence \citep{iskakov2007numerical,schekochihin2007fluctuation}. It would reasonable that they are of comparable magnitudes $Rb_{m,M}^c\sim Rb_{m}^c$ since both act as effective outer magnetic Reynolds numbers for their respective isotropic subranges.

\section{Conclusion}
Critical balance is a theory for strong turbulence in anisotropic wave systems that argues for a balance of linear wave and non-linear interaction timescales to predict the scale-by-scale structure of the turbulent cascade. We have proposed that critical balance can unify the unity $Pr$ geophysical regime and the extremely low $Pr$ astrophysical regime of strongly stratified turbulence because both support anisotropic linear wave motions in different asymptotic limits of the internal gravity wave dispersion relation. The dispersion reduces to adiabatic, inviscid IGWs in the $Pr=O(1)$ limit and IGWs overdamped on buoyancy-modified timescales in the $Pr\ll1$ limit. We find that a smooth transition between the two regimes occurs at $PrRb=O(1)$ as $Pr$ is decreased, or equivalently when the turbulent Peclet number $Pe_t=u_zl_z/\kappa$ drops below order unity and shifts dominant balance in the buoyancy equation. Application of critical balance in the $PrRb<1$ regime predicts an anisotropic cascade and scaling relations for the outer vertical scales that are identical to the geophysical regime if the Froude number, $Fr$, is simply replaced by a modified Froude number $Fr_M\equiv Fr/(PrRb)^{1/4}$ (e.g. the outer vertical length scale is $l_z\sim Fr_ML$). Indeed, a smooth transition occurs at $PrRb=1$ between the two regimes.

The scaling relations from critical balance offer a theoretical framework for understanding properties of strongly stratified turbulence in stellar radiative zones. Application to estimating the vertical turbulent diffusion coefficient in the $PrRb<1$ regime gives the same scaling result as the original estimates of \cite{zahn1992circulation}, thereby their validating their dimensional arguments. However, a complete understanding of strongly stratified turbulence in stellar radiative zones would need to also take into account the effects of non-local energy transfer, rotation, and magnetic fields (large and small scale). In analogy with the geophysical regime, energy at large scales in the thermally diffusive regime is likely also transferred through non-local transfer mechanisms, not only through a local energy cascade as assumed in our critical balance arguments. Rotation may have a significant effect when the driving horizontal scales  are sufficiently large to be influenced by Coriolis forces, which is thought to be the case in the upper solar radiative zone \citep{garaud2020horizontal}, for example. Similarly, large scale magnetic fields will modify the wave dispersion relation and support magneto-internal waves, whose properties will change the nature of the turbulence presented here depending on the magnitude and direction of the large scale field. Lastly, small scale magnetic fields can grow in the isotropic portion of the turbulence through the small-scale dynamo if the stratification is not too strong in the $PrRb>1$ regime \citep{skoutnev2021small}. We proposed a modified instability criterion for the SSD in the $PrRb<1$ limit based on the new scaling laws. The modified instability criterion predicts a more easily excited dynamo since thermal diffusion ameliorates the effect of stratification. Stellar stratified turbulence is therefore, at the very least, in a saturated state of the SSD, containing small-scale magnetic fields in near equiparition with the isotropic scales of the velocity field. The effects of a saturated SSD on mixing and transport are unknown.

\medskip
The author would like to thank Matthew Kunz for an early suggestion to explore more generally the ideas of critical balance, Amitava Bhattacharjee for providing intuition of their application to MHD turbulence, and Pascale Garaud for many insightful discussions on stratified turbulence that eventually led to this work. We also thank the referees for improvements to the physical interpretation of the results. V. S. was supported by Max-Planck/Princeton Center for Plasma Physics (NSF grant PHY-1804048). We thank Kailey Whitman  for illustration of the graphical abstract.  Declaration of Interests. The authors report no conflict of interest.

\bibliographystyle{jfm}
\bibliography{references}

\end{document}